\documentclass[twocolumn,showpacs,preprintnumbers,amsmath,amssymb,aps,prb]{revtex4-1}
\usepackage{graphicx}

\begin{document}
\title{
Conductance through disordered graphene nanoribbons: Standard and anomalous electron localization
}
\author{Ioannis Kleftogiannis}
\author{Ilias Amanatidis}
\affiliation{Department of Physics, University of Ioannina, Ioannina 45110, Greece}
\author{V\'ictor  A. Gopar }
\affiliation{Departamento de F\'isica Te\'orica and BIFI, Universidad de Zaragoza, Pedro Cerbuna 12, E-50009, Zaragoza, Spain.}

\begin{abstract}
Conductance fluctuations produced by the presence of disorder in zigzag and armchair graphene nanoribbons are studied. We show that quantum transport in zigzag nanoribbons takes place via edge states which are exponentially localized, as in the standard Anderson localization problem, whereas for  armchair nanoribbons  the symmetry of the graphene sublattices produces anomalous localization, or delocalization. We show that these 
two different electron localizations lead to significant differences of  the conductance statistics  between  zigzag and armchair nanoribbons. In particular,  armchair nanoribbons show nonconventional large conductance fluctuations relative to those of  Anderson-localized electrons.   We calculate analytically the complete  distribution of conductances for both types of ribbons.  Without free fitting parameters, we verify our theoretical results by performing  numerical simulations  of disordered zigzag and armchair nanoribbons of experimentally achievable lengths and widths.

\end{abstract}
\pacs{72.80.Vp, 72.10.-d,73.20.Fz, 73.20.Jc}
\maketitle

\section{introduction}
Graphene remains  one of the most studied materials since the experimental isolation of its  two-dimensional honeycomb lattice from graphite.\cite{novoselov2004}  The singular properties of  graphene have attracted such an interest, 
 from fundamental and application points of view, that  nowadays an extensive  literature exists on different physical phenomena occurring in this material. \cite{reviews} The honeycomb arrangement of carbon atoms in graphene provides two relevant  ingredients to  electronic transport: a sublattice (chiral) symmetry 
 and a linear  dispersion relation.  The latter, valid at low energies,\cite{wallace}  has attracted a 
 lot of attention since it  allows  an  analogy with  relativistic massless  particles  described by the Dirac equation.  However, the sublattice symmetry is perhaps a  feature  of more fundamental importance. The lattice structure of  graphene is described by two identical interconnected sublattices, such that  the sites of one sublattice are connected
only to the sites of the other one, i.e., the lattice of graphene is bipartite resulting in a symmetric  energy spectrum around the Fermi energy. The sublattice symmetry,  also known as chiral symmetry,  has deep consequences on different  electronic properties, as  has been shown in the 
past, \cite{soukoulis,inui,spiros1}  although,  there is a revived interest in the context of superconducting quasiparticles, topological insulators,  and superconductors. \cite{altland,schnyder,xiao}
 
 Electronic properties of pristine graphene structures have been extensively studied. However,  real graphene-based  devices can be affected by the presence of disorder, as  any other mesoscopic material. Different  sources of disorder can be found in graphene. For instance, distortion of 
 the lattice like ripples can appear by interactions with the substrate. \cite{gallagher,melinda} Experiments have been performed using 
 suspended graphene structures to avoid effects of the substrate, \cite{meyer,jarillo} 
but even in this case graphene samples  are not free of defects. 
Also, experimental evidence of  short range disorder in graphene has been found in Refs. \onlinecite{Tan,Jang}. Despite the fact that disorder is generally considered as an ingredient to be avoided, one 
might  take advantage of it. For instance, electronic properties of graphene can be manipulated by doping a sample or by replacing carbon atoms.  \cite{pi, ruitao, peres}

Disorder plays a central role in the problem of quantum transport and therefore its effects have been widely studied.  For instance, it is known that  in one-dimensional (1D) systems,   the presence of any small amount of disorder leads to  exponential localization of electron wavefunctions (Anderson localization) with the distance $r$:  $| \psi|  \sim e^{-\gamma r}$. This exponential decay  has 
been experimentally and theoretically studied in different  disordered systems.\cite{phystoday}  The  
inverse of the localization length $\gamma$ is of fundamental importance for the description of statistical properties of transport within the 
scaling approach to localization.\cite{mello_book}  However, electron wavefunctions are not always exponentially (Anderson) localized; it has been shown that the wavefunction decays  as  $| \psi|  \sim e^{-\gamma \sqrt{r}}$ at the band center of 
 1D  systems with random hopping between neighboring lattice sites (off-diagonal disorder), i.e.,  electrons are less localized  (delocalized) than in the standard  Anderson problem.\cite{soukoulis,spiros1}

Several  models have been used to describe different effects of disorder in graphene. The literature is very extensive and 
we  turn the attention of the  reader to  Ref. \onlinecite{mucciolo} for a review of the topic (see also Refs. \onlinecite{areshkin, bardarson,kentaro, pablo,rossi,reviews,xiong,verges,evaldsson,libisch,wakabashi,ioannis_1,ioannis_2}). For example,  effective Hamiltonians (Dirac Hamiltonians) valid in the continuum low-energy limit and  tight-binding  models have been introduced to study statistical properties of conductance fluctuations, such as the density of transmission eigenvalues.  
\cite{areshkin, bardarson,kentaro, pablo,rossi,reviews,xiong,verges,evaldsson,libisch,wakabashi}   Most of the theoretical works, however,  deal with effects produced by  the standard Anderson-localization of electrons.

For finite-size graphene structures like  nanoribbons, an additional  element has to be considered in the study of quantum transport,  namely, the sample terminations or edges. There are  two different basic graphene edges: zigzag and armchair,  which 
 affect the electronic properties of nanoribbons. For example, the band structure is  completely different for each  
 termination (e.g.,  Fig. \ref{figure_1}, bottom  panels). Also, the existence of edge states in clean zigzag nanoribbons has been shown, \cite{nakada,tao}  whereas in armchair nanoribbons those states are absent. 
 
 Detailed experimental  investigations on the effects  of graphene edges on quantum transport  have been restricted  by technological limitations on controlling the edge of graphene samples. Nowadays  a precise manipulation 
 of graphene terminations is still a  challenge, but  recent experiments have shown 
a high control of the nanoribbon edges.\cite{zhang} In fact, transport experiments have been performed with an accurate control of the  nanoribbon terminations. \cite{koch} 

Therefore, understanding the role of the edge morphology in  nanoribbons as well as the effects of disorder in 
graphene-based devices is of interest from a fundamental point of view and relevant to  future experimental investigations.

In this work,  the effects of both  disorder and edges on the conductance fluctuations of graphene nanoribbons are studied. We 
calculate the conductance distribution of disordered zigzag and armchair graphene  nanoribbons.  
As we show below,  electronic transport in zigzag nanoribbons takes place via  standard exponentially localized states (Anderson localization), which allows us  to study the conductance statistics within a conventional approach to localization. For armchair nanoribbons, however,  electrons are delocalized and we require an extension of the standard localization approach. Thus,  the conductance statistics is  affected significatively by the nanoribbon edges.

\section{Numerical and theoretical models}
Concerning the numerical results, we use a standard tight-binding Hamiltonian model to describe the armchair and zigzag graphene nanoribbons:
\begin{equation}
\label{tight}
  H= \sum_{\langle i,j \rangle}t_{i,j} (c_{i}^\dagger c_{j} +c_{j}^\dagger c_{i} ) , 
\end{equation}
where $i$ and $j$ are nearest neighbors and $c_i^\dagger$ ($c_i$) is the creation (annihilation) operator for spinless fermions. Disorder is 
introduced via   random hopping elements $t_{i,j}$ connecting the two sublattices. This type of short-range disorder models the presence of distortions in graphene samples and preserves the symmetry of the graphene lattice. The $t_{i,j}$'s  
are sampled from the distribution $p(t)=1/wt$ with $\exp(-w/2) \leq t \leq \exp (w/2) $, where $w$ denotes the strength of the disorder (we fix  $w=1$). 
Our numerical calculations are performed  at the low  energy  $E=10^{-6}$ (in units of the hopping energy of the perfect leads).  The 
length $L$ and width $W$ of the nanoribbons are reported in units of the lattice constant $a \simeq 2.46 \textup{\AA}$. We have fixed  
$W=11a/ \sqrt{3}$ for zigzag nanoribbons in our simulations, which corresponds to eight  zigzag chains, while for armchair ribbons $W=5 a$. 
The conductance $G$ (in units of the conductance quantum $ {2e^{2}}/{h}$)  is calculated  by attaching perfect metallic graphene leads
to the left and right side of the disordered ribbons  (top panels in Fig. \ref{figure_1}). We use a recursive Green's function 
method \cite{li} to calculate the conductance within the Landauer-B\"uttiker approach. The conductance statistics is obtained by collecting data over an ensemble of $2 \times 10^4$  disorder realizations. 

\subsection {Disordered zigzag-nanoribbons} 
\begin{figure}
\includegraphics[width=\columnwidth]{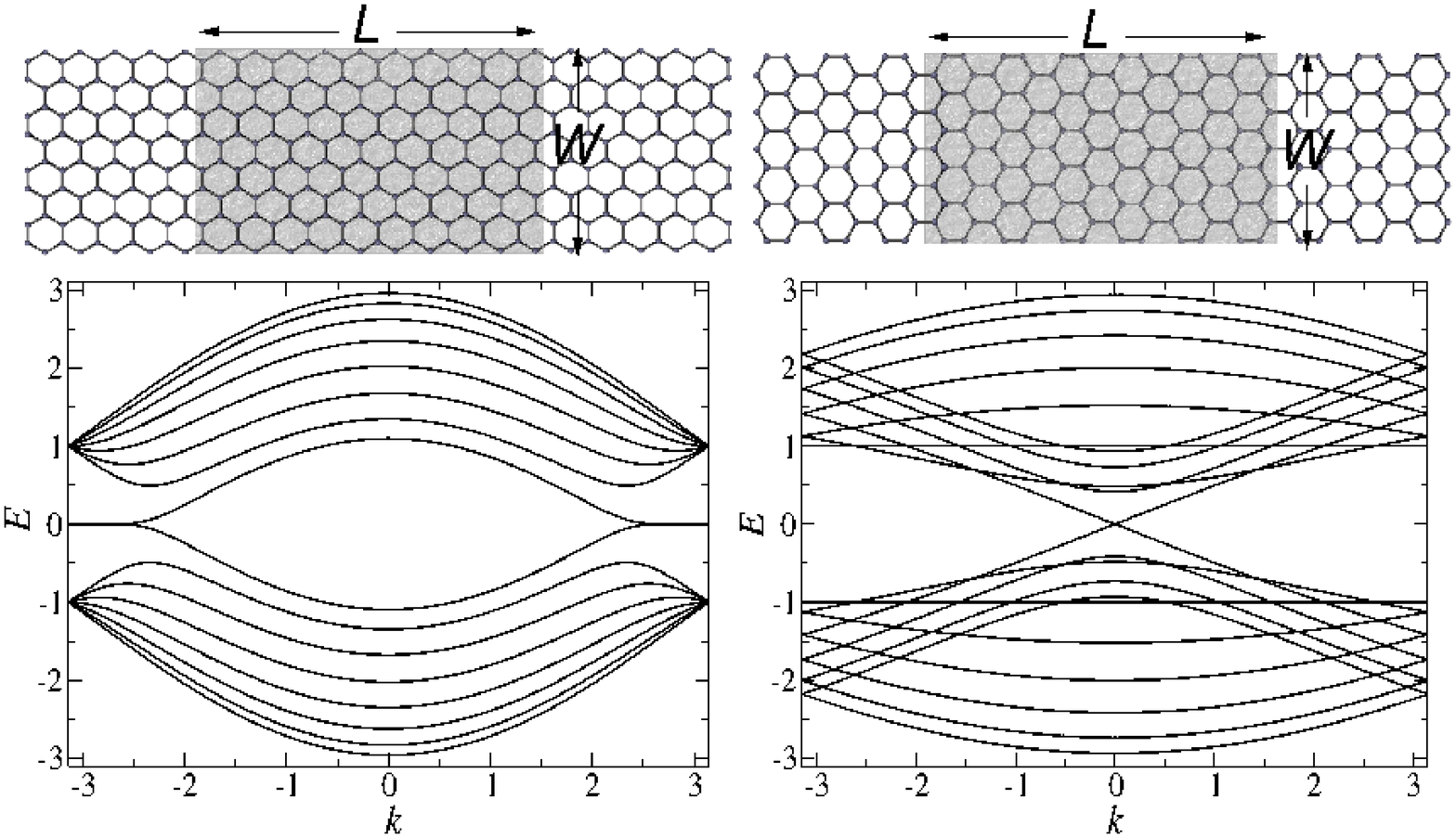}
  \caption{Top: Graphene nanoribbons with zigzag (left) and armchair (right) edges with disordered regions  (shaded areas)  of length $L$ and width $W$ attached to perfect leads. Bottom:   Band structures calculated for the clean  zigzag (left) and armchair (right) graphene nanoribbons of width $W=11 a/\sqrt{3}$ and 5$a$, respectively.}
  \label{figure_1}
  \end{figure}
\begin{figure}
\includegraphics[width=\columnwidth]{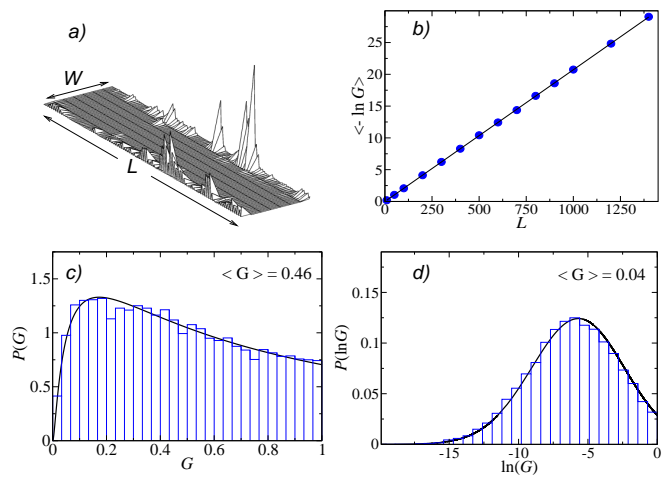}
  \caption{(Color online) (a) Modulus squared of the wavefunction  of a zigzag nanoribbon sample. We can see that the wavefunction is concentrated along the  border of the sample. According to the  conductance fluctuations these edge states are exponentially (Anderson) localized. (b)  The linear dependence of $\langle \ln G \rangle$ with $L$ for zigzag nanoribbons. The solid line is a linear fitting of the numerical data (dots). (c) Numerical (histogram) and theoretical (solid line) conductance distributions for $\langle G \rangle =0.46$  ($\langle - \ln G \rangle =1.03$).  (d) Numerical (histogram) and theoretical (solid line) distributions of the logarithm of the conductance for the case  $\langle G \rangle =0.04$ ($\langle - \ln G \rangle =6.21$). A good agreement is seen between  numerical and theoretical distributions, in both panels (c) and (d).}
  \label{figure_2}
 \end{figure}

From the band structure of  pristine zigzag nanoribbons  (bottom-left panel in Fig. \ref{figure_1}), we see  that only one 
channel is open at low energies ($E \lesssim 0.5$). On the other hand, it is known that edge states are 
present in perfect graphene samples with zigzag terminations.\cite{nakada} When disorder is introduced, those states living at the border of the 
nanoribbon become exponentially localized. As an illustration of those edge localized states, in Fig. \ref{figure_2} (a) we show the modulus squared of the wavefunction  for one realization of the disorder; we 
can see that the wavefunction is located on the border slightly penetrating the ribbon. In  the Anderson localization problem, it is known that  the conductance decays exponentially with the length $L$ and a signature of (Anderson) localization is the linear behavior of the average $\langle \ln G \rangle$ with  $L$. We have obtained numerically such a linear relation, as  can be seen in Fig. \ref{figure_2} (b).  

The exponential localization of electrons for zigzag nanoribbons simplifies  our analysis  since, within a scaling theory of Anderson localization,  the conductance distribution $P_s (G)$ is determined completely by the ratio $s\equiv L/l$, $l$  being the mean free path. \cite{mello_book,melnikov,dorokhov}  The exact  expression for $P_s (G)$ is given in terms of quadratures. Here we provide a  simpler expression:  
\begin{equation}
\label{pofG}
 P_s(G)=C  \frac{\left[\mathrm{acosh}(1/\sqrt{G})\right]^{1/2}}{G^{3/2}(1-G)^{1/4}}e^{-(1/s)\mathrm{acosh}^2(1/\sqrt{G})} ,
\end{equation}
where $C$ is a normalization constant. Equation  (\ref{pofG}) is an approximation to  the exact solution of Mel'nikov equation (see the Appendix) 
and   is useful for any practical value of $s$. The value of $s$ can be determined numerically through the relation  $\langle -\ln G \rangle=L/l$. \cite{gopar-molina}

We now compare some results from our expression Eq. (\ref{pofG}) with numerical simulations of zigzag nanoribbons. In Fig. \ref{figure_2} (bottom) we show two  numerical distributions (histograms) and the  corresponding theoretical predictions (solid lines).  
The value of $s$ in  Eq. (\ref{pofG}) is extracted from our  simulations [Fig. \ref{figure_2}](b)]. We present two cases with very 
different average conductance: for Figs. \ref{figure_2}(c) and (d), $\langle G \rangle $=0.46 and  $\langle G \rangle = 0.04$, respectively.
Note that, 
while in Fig. \ref{figure_2}(c) we show $P_s(G)$, in Fig. \ref{figure_2} (d) $P_s(\ln G)$ is plotted instead, as 
for small values of $\langle G \rangle$, the distribution $P_s(G)$ is  very sharp. An excellent agreement between simulations and theory  is seen in both panels (c) and (d) in  Fig. \ref{figure_2}.  
 
\subsection{Disordered armchair nanoribbons} 
\begin{figure}
\includegraphics[width=\columnwidth]{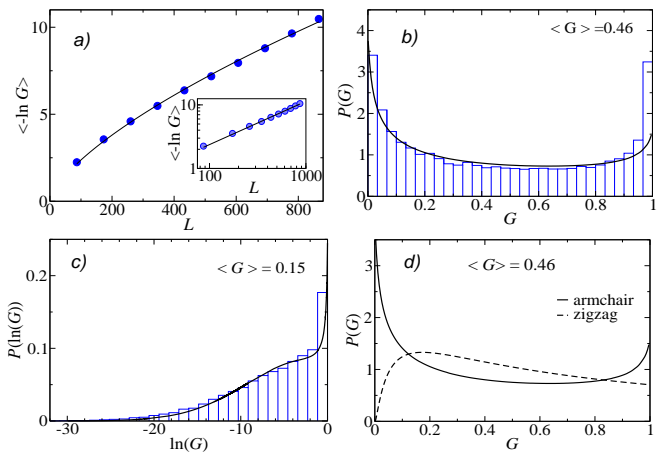}
  \caption{Color online (a) The nonlinear dependence of the average $\langle \ln G \rangle$ with  $L$ for  armchair nanoribbons. The solid line is obtained by fitting: $\langle \ln G \rangle \propto L^\alpha$ with $\alpha=0.69$. Inset: 
  the  power-law dependence is valid at all nanoribbon lengths, as shown by  the single straight line obtained when plotting the data in log-log 
  scale.  (b) Numerical (histogram) and theoretical (solid line) distributions for armchair ribbons with $\langle G \rangle =0.46$ ($\langle -\ln G = 1.37 \rangle$). (c) Numerical (histogram) and theoretical (solid line) distributions $P(\ln G)$ for armchair ribbons with $\langle G \rangle =0.15$ 
  ($\langle -\ln G \rangle = 6.37 $).  In both panels (b) and (c)   a good agreement is seen. (d) 
  Comparison of the conductance distributions  for armchair (solid line)  and zigzag (dashed line) nanoribbons with the same average $\langle G \rangle =0.46$.
  }
  \label{figure_3}
  \end{figure}

We  observe from the band structure in Fig. \ref{figure_1} (bottom-right) that only one channel contributes to the transport near the Fermi energy, as in the  case of zigzag nanoribbons. 
Now, however,  anomalous localization of electrons takes place.  A signature of  such delocalization is the nonlinear dependence of $\langle \ln G \rangle$ versus the  length $L$. We illustrate this behavior in our armchair nanoribbons  in Fig. \ref{figure_3}(a), where a  power-law dependence of $\langle \ln G \rangle$ with  $L$ is shown. One 
might contrast this result with the  linear dependence of $\langle \ln G \rangle$ with $L$  for zigzag nanoribbons [Fig. \ref{figure_2}(b)]. 

Therefore anomalous localization is present in our armchair nanoribbons and we  cannot study the electronic transport within the standard scaling approach of localization, Eq. (\ref{pofG}). 

A  model  to describe the statistical  properties of transport when electron wavefunctions are anomalously localized
has been proposed.\cite{fernando_gopar, elias}  Within that  model,  the knowledge of just two quantities:  the average $\langle \ln G \rangle$ and the exponent $\alpha$ in $\langle \ln G \rangle \propto L^\alpha$,  is enough  to calculate the complete  distribution of conductances. Physically, $\alpha$ might be considered  a  measure of the strength of the electron localization: the localization becomes  weaker as $\alpha$ 
decreases.  With  the above information, the conductance distribution $P_{\xi, \alpha}(G)$ can be obtained from  
\begin{eqnarray}
\label{pofG_xi}
P_{\xi,\alpha}(G)=\int_0^\infty P_{s(\alpha,\xi,z)}(g) q_{\alpha,1}(z){\rm d}z  ,
\end{eqnarray} 
where we have defined $\xi =\langle \ln G \rangle$. $P_{s(\alpha,\xi,z)}(G)$ is given by  Eq. (\ref{pofG}), $s$ being now a function defined as $ s(\alpha,\xi,z)={\xi}/(2{z^\alpha I_\alpha)}$, where $I_\alpha =1/2 \int_{0}^{\infty} z^{-\alpha} q_{\alpha,1}dz $. The function  
$q_{\alpha,1}(z)$ is the probability density of the so-called $\alpha$-stable distributions ($\alpha <1$) supported in the positive semiaxis.  No analytical expressions for $\alpha$-stable distributions exist, except for few special values of $\alpha$. We thus obtain  $P_{\xi,\alpha}(G)$  by numerical integration.

As we pointed out, our theoretical model depends on   $\alpha$ and   $\langle \ln G \rangle$, which are extracted from the numerical simulations and then plugged  into Eq. (\ref{pofG_xi}). From Fig.  \ref{figure_3}  we have extracted the value of $\alpha=0.69$. This value does not depend on the length, width, and degree of disorder, according to our numerical simulations; in this sense it is universal for graphene 
nanoribbons with off-diagonal disorder, at the band center.\cite{alpha1D}  Thus, 
in Figs. \ref{figure_3} (b) and \ref{figure_3}(c) we compare the numerical (histograms) and theoretical (solid lines) distributions  for two different conductance averages.  In Fig. \ref{figure_3}(b)  we show $P(G)$ for a case with $\langle G \rangle = 0.46$, while in Fig.   \ref{figure_3}(c), 
$\langle G \rangle = 0.15$. Note that in the latter case we plot $P(\ln G)$ for convenience, as we previously explained.  Thus, we see that our model describes the trend of the complete conductance distributions.

In order to remark the differences in the conductance fluctuations between zigzag and armchair disordered nanoribbons, we  plot in Fig. \ref{figure_3}(d)  the distributions for both edge terminations with the same conductance average $\langle G \rangle=0.46$. 
We  see  that  distributions are very different.  We notice, for instance,  that  small conductances ($G \sim 0$) are suppressed for zigzag nanoribbons, in contrast to the case of armchair nanoribbons whose  conductance distribution has a peak. In addition, large values of conductance ($G \sim 1$) are favored in armchair nanoribbons, in relation to the zigzag case. 

\section{Summary and discussion}
We  have studied the quantum conductance fluctuations of disordered zigzag and armchair nanoribbons. For both ribbon edges,  we provide  analytical expressions of the conductance distributions which are  determined by the average $\langle \ln G \rangle$
and the exponent $\alpha$ of its power-law decay with the system length. To our knowledge, there is no other theoretical framework that provides the complete conductance distribution. 
Without free parameters, the  theoretical predictions have been confirmed by numerical simulations of graphene nanoribbons. The simulations have been performed using a tight-binding  model with random hopping elements which preserves the chiral symmetry on  the  graphene lattice. 

Our results show significant differences between the conductance statistics of armchair and zigzag nanoribbons. Essentially,  this is a consequence of the different way  that electrons become localized in each type of nanoribbon.  

We have found  that electrons in zigzag ribbons  
are more  localized than in armchair ribbons:  while in zigzag nanoribbons electrons are exponentially localized, as in the standard Anderson localization, in armchair nanoribbons electrons are anomalously localized (delocalized). The strength of delocalization 
(measured by $\alpha$) does not depend on the length, width, or disorder degree of the armchair ribbon, it rather  being fundamentally   an 
effect of the chirality of the honeycomb lattice. Electronic transport in zigzag nanoribbons 
takes place via states located at the border of the ribbons which correspond to a flat band at low energies (Fig. \ref{figure_1} bottom left).  We  have shown that  under the 
presence of disorder, those edge states become exponentially localized. In contrast, electrons  in armchair nanoribbons  follow a linear 
dispersion relation  (Fig. \ref{figure_1}, bottom right),  as in other disordered systems that exhibit  delocalization effects (e.g.,  Refs.  \onlinecite{brouwer,steiner}).  

We remark that our  model is valid for 1D systems (one channel) while nanoribbons are 2D structures. The metallic bands 
(Fig. \ref{figure_1}) show that  transport occurs effectively through one single channel, near the Fermi energy. On the other hand, first-principle calculations have shown that band structures of armchair and zigzag nanoribbons always have an energy gap.\cite{son,gunlycke}  The band structures, however,  become similar to those considered here for slightly wider nanoribbons, near the band gap. We thus expect  to observe similar localization/delocalization effects to those studied here.

Concerning the experimental observability of our results, we point out that the effects of delocalization on the conductance statistics  depend  on the conservation of the chiral symmetry of the graphene sublattices. Chiral symmetry can be broken by different sources; for instance, 
by  the presence of  lattice deformations like the so-called Stone-Wales defects,\cite{stone-wales,ma, ibolya}  which lead  to the formation of pentagons and heptagons on the honeycomb lattice of graphene.  The  
presence of this kind of defect  in armchair ribbons might prevent delocalization of electrons, but  delocalization effects would be  gradually 
washed out  as  the number of those defects are increased and, eventually, standard Anderson localization would take place. Another  issue  is the  
effect of  Coulomb interactions,\cite{valeri} which we have neglected.  Coulomb interactions 
break the chiral symmetry while they  become more important 
in suspended graphene samples, since no substrate can screen the electromagnetic fields.  Recent calculations, however,  have shown the restoration of the chiral symmetry by  lattice distortions, like the  so-called Kekul\'e distortion.\cite{araki}

To conclude, we have assumed well defined nanoribbon edges, zigzag or armchair,   which  highlights the relevance of edges to electronic transport.  Nanoribbons with mixed edges can be analyzed within our framework; in this case, we expect to observe  intermediate conductance fluctuations between those studied here. Nonetheless, the observability of our findings depends on the control of the nanoribbon edges. Recent 
advances \cite{lijie,koch, zhang} indicate that in the near future experiments will be performed allowing  full control of the nanoribbon 
terminations. Thus,  we hope  that our 
results motivate further experimental investigations and lead to a better understanding of the central role that disorder and symmetries play  on  quantum electronic transport through graphene  nanoribbons.

\begin{acknowledgements}
We thank A.  Cruz for carefully reading  our manuscript.  
We acknowledge  support from  MINECO (Spain) under Project  No. FIS2012-35719-C02-02 and  the  
HPC-EUROPA2 project (Project No. 228398) with the support of the European Commission-Capacities Area-Research Infrastructures. 
\end{acknowledgements}

\appendix*
\section{Derivation of Eq. (\ref{pofG})}

Within a scaling approach to localization, \cite{melnikov, dorokhov, mello_book}  the distribution of the dimensionless conductance $P_s(G)$ is determined by the solution of a Fokker-Planck equation (Mel'nikov equation), which is an evolution equation for  $P_s(G)$ as a function of the length $L$ of the disordered system. For one channel, such an evolution equation is usually written in terms of the variable $\lambda$ as  
\begin{equation}
\label{melnikov}
 \frac{\partial p_s(\lambda)}{\partial s}=\frac{\partial}{\partial \lambda}\left[\lambda(1+\lambda) \frac{\lambda p_s(\lambda)}{\partial \lambda}  \right] ,
\end{equation}
where $s=L/l$,  $l$ being the mean free path. The variable $\lambda$ $(0 < \lambda < \infty)$ is related to the conductance by
\begin{equation}
\label{lambda}
\lambda=({1-G})/{G} . 
\end{equation}
The solution to the differential equation (\ref{melnikov}) is known in terms of quadratures. \cite{abrikosov}  We notice that the distribution of $\lambda$, and therefore the distribution of $G$,  depends only on the parameter $s$.

For convenience,  we rewrite Eq. (\ref{melnikov}) in terms of the variable $x$,  where  $\lambda=\sinh^2 x$. After some algebra we obtain  
\begin{equation}
 \frac{\partial p_s(x)}{\partial s}=\frac{1}{4}\frac{\partial}{\partial x}\left[  \frac{\partial p_s(x)}{\partial x}  -2 \coth(2x) p_s(x)    \right]  .
\end{equation}
The solution $p_s(x)$ of this partial differential equation is given by 
\begin{equation}
\label{pofx}
p_s(x)= \frac{1}{\sqrt{\pi}}\left( \frac{1}{2 s}\right)^{3/2} e^{-s/4} \sinh(2x) I(x) ,
\end{equation}
where we have defined 
\begin{equation}
\label{Iofx}
 I(x)=\int_x^\infty dy \frac{y e^{y^2/s}}{\sqrt{\cosh(2y) -\cosh(2x)}} .
\end{equation}

Now, the main contribution to the integral $I(x)$ comes from values $y \approx x$.  We thus expand the integrand around $y=x$:
\begin{eqnarray}
 \frac{y e^{y^2/s}}{\sqrt{\cosh(2y) -\cosh(2x)}} \approx& & \frac{1}{\sqrt{2}}\frac{x e^{-x^2/s}}{\sqrt{\sinh(2x)}}\frac{1}{\sqrt{y-x}} \nonumber \\
 &+& \mathcal{O}[\sqrt{y-x}] .
\end{eqnarray}
Plugin the first term of this expansion to  $I(x)$, Eq. (\ref{Iofx}), and restricting the upper limit of  the integral to keep valid our first 
order approximation, we find  that
\begin{equation}
\label{Iofxapprox}
 I(x)\approx \sqrt{\frac{x}{2 \sinh(2x)}} e^{-x^2/s}.
\end{equation}
Thus, substituting Eq. (\ref{Iofxapprox}) into Eq. (\ref{pofx}), we obtain
\begin{equation}
\label{pofxapprox}
 p_s(x)=C_s \sqrt{x \sinh(2x)}e^{-x^2/s} ,
\end{equation}
where we have defined the normalization constant $C_s$, which depends on the parameter $s$. 

We now write our expression for  $p_s(x)$, Eq. (\ref{pofxapprox}),  in terms of the conductance $G$. From Eq. (\ref{lambda}) and using that $\lambda=\sinh^2x$, we have that  $G=1/\cosh^2x$.  The distribution of conductances is thus given by $P_s(G)=\left| dx/dG \right| p_s(x)$. 
After some algebraic  simplifications, we finally obtain  
\begin{equation}
 P_s(G)=C  \frac{\left[\mathrm{acosh}(1/\sqrt{G})\right]^{1/2}}{G^{3/2}(1-G)^{1/4}}e^{-(1/s)\mathrm{acosh}^2(1/\sqrt{G})} ,
\end{equation}
which is the expression for the conductance distribution  shown in the main text.


\begin{thebibliography}{60}

\bibitem{novoselov2004} K. S. Novoselov, A. K. Geim, S. V. Morozov, D. Jiang, Y. Zhang, S. V. Dubonos, I. V. Grigorieva, A. A. Firsov, Sience {\bf 306}, 666, (2004).

\bibitem{reviews} We refer the reader to the recent review papers:  A. Castro Neto, F. Guinea, N. Peres, K. Novoselov, and  A. Geim, Rev. Mod. Phys. {\bf 81}, 109 (2009); 
S. Das Sarma, Shaffique Adam, E. Hwang, and Enrico Rossi, Rev.  Mod. Phys.  {\bf 83}, 407 (2011).

\bibitem{wallace} P. R. Wallace, Phys. Rev. {\bf 71}, 622 (1947).

\bibitem{soukoulis} C. M. Soukoulis and E. N. Economou, Phys. Rev. B {\bf 24}, 5698  (1981).

\bibitem{inui}  M. Inui, S. A. Trugman, Elihu Abrahams Phys. Rev. B {\bf 49}, 3190 (1994).

\bibitem{spiros1} S. N. Evangelou and D. E. Katsanos, J. Phys. A {\bf 36}, 3237 (2003).

\bibitem{altland}A. Altland and M. R. Zirnbauer, Phys. Rev. B {\bf 55}, 1142  (1997). 

\bibitem{schnyder} A.  P. Schnyder, S.  Ryu, A.  Furusaki, and A. W.  W. Ludwig, Phys. Rev. B {\bf 78}, 195125 (2008).

\bibitem{xiao} X.-L. Qi and S.-C. Zhang, Rev. Mod. Phys. {\bf 83}, 1057 (2011).

\bibitem{gallagher}
P. Gallagher, K. Todd, and D. Goldhaber-Gordon, Phys. Rev. B {\bf 81}, 115409 (2010).

\bibitem{melinda} 
M. Y. Han, J. C. Brant, and P. Kim, Phys. Rev. Lett, {\bf 104}, 056801, (2010).

\bibitem{meyer} 
J. C. Meyer, A. K. Geim, M. I. Katsnelson K. S. Novoselov, T. J. Booth, and S. Roth,  Nat.  Mater. {\bf 10}, 282 (2011).

\bibitem{jarillo} J. Xue, J. Sanchez-Yamagishi,  D. Bulmash, P.  Jacquod, A. Deshpande, K. Watanabe, T.  Taniguchi, P.  Jarillo-Herrero, B. J.  LeRoy, Nature Mat. {\bf 10} , 282 (2011).


\bibitem{Tan}Y.-W. Tan, Y. Zhang, K. Bolotin, Y. Zhao, S. Adam, E.-H. Hwang, S. Das Sarma, H.-L. Stormer, and P. Kim, Phys. Rev. Lett. {\bf 99}, 246803 (2007);

\bibitem{Jang}C. Jang, S. Adam, J.-H. Chen, E. D. Williams, S. Das Sarma, and M. S. Fuhrer, Phys. Rev. Lett. {\bf 101}, 146805 (2008).

\bibitem{pi} K. Pi, K. M. McCreary, W. Bao, Wei Han, Y. F. Chiang, Yan Li, S.-W. Tsai, C. N. Lau, and R. K. Kawakami
Phys. Rev. B {\bf 80},  075406 (2009).

\bibitem{ruitao} R. Lv and M. Terrones, Mater. Lett., {\bf 78}, 209, (2012).

\bibitem{peres} N. M. R. Peres, L. Yang, and S.-W. Tsai, New J  Phys., {\bf 11}, 09500, (2009).

\bibitem{phystoday} A.  Lagendijk, B. van Tiggelen, and D.  S. Wiersma, Phys. Today {\bf 62}(8), 24, (2009), and references therein.

\bibitem{mello_book} P.  A. Mello and N. Kumar, {\it Quantum Transport in Mesoscopic Systems}, (Oxford University Press, Oxford, 2004).

\bibitem{mucciolo} E. R. Mucciolo and C. H. Lewenkopf, J.  Phys.: Condens Matter {\bf 22}, 273201 (2010).

\bibitem{areshkin} D. A. Areshkin, D. Gunlycke, and C. T. White,  Nano. Lett. {\bf 7}, 204 (2007).

\bibitem{bardarson}  J. H. Bardarson, J. Tworzydlo, P. W. Brouwer, and C. W.  J.  Beenakker,  Phys. Rev. Lett. {\bf 99}, 106801 (2007).

\bibitem{kentaro} K.  Nomura, M. Koshino, and S. Ryu , Phys. Rev. Lett.  {\bf 99}, 146806 (2007) .

\bibitem{pablo}P. San-Jose, E. Prada, D. S. Golubev,  Phys. Rev. B {\bf 76}, 195445 (2007). 

\bibitem{rossi} E. Rossi, J. H. Bardarson, M. S. Fuhrer, and S. Das Sarma, Phys. Rev. Lett. {\bf 109}, 096801 (2012).
 
\bibitem{xiong} Shi-Jie Xiong and Ye Xiong,  Phys. Rev. B {\bf 76}, 214204 (2007).

\bibitem{verges} E. Louis, J. A. Verg\'es, F. Guinea, and G. Chiappe, Phys. Rev. B, {\bf 75}, 085440 (2007).

\bibitem{evaldsson} M. Evaldsson, I. V. Zozoulenko, H. Xu, and T. Heinzel, Phys. Rev. B {\bf 78} 161407(R) (2008).

\bibitem{libisch} F. Libisch, S. Rotter, and J. Burgd\"orfer, New J. Phys. {\bf 14}, 123006  (2012).

\bibitem{wakabashi} K. Wakabayashi, Y. Takane, and M. Sigrist, Phys. Rev. Lett. {\bf 99}, 036601 (2007).

\bibitem{ioannis_1}
I. Kleftogiannis, S. N. Evangelou, arXiv:1304.5968.

\bibitem{ioannis_2}
H. Amanatidis, I. Kleftogiannis, D.E. Katsanos, S.N. Evangelou,  arXiv:1302.2470.

\bibitem{nakada} 
K.  Nakada,  M.  Fujita, G. Dresselhaus, and M.  S. Dresselhaus, Phys. Rev. B {\bf 54}, 17954 (1996).

\bibitem{tao}
C. Tao, L. Jiao, O. V. Yazyev, Y.-C. Chen, J. Feng, X. Zhang, R. B. Capaz, J. M. Tour, A. Zettl, S. G. Louie, H. Dai, and M. F. Crommie, Nat. Phys. {\bf 7}, 616 (2011).

 \bibitem{zhang} 
 X. Zhang, O. V. Yazyev, J.  Feng,  L. Xie, C.  Tao, Y.-C. Chen, L.  Jiao, Z.  Pedramrazi, A.  Zettl, S.  G. Louie, H.  Dai, M.  F. Crommie, 
 ACS Nano {\bf 7}, 198 (2013). 
 
\bibitem{koch} 
M. Koch, F. Ample, C. Joachim, and L. Grill, Nat.  Nanotechnol. {\bf 7}, 713 (2012).
 
 \bibitem{li}
 T. Li, Q. W. Shi, X. Wang, Q. Chen, J. Hou, and J. Chen, Phys. Rev. B {\bf 72}, 035422 (2005).
 
 \bibitem{melnikov} 
V. I.  Mel'nikov JETP Lett., {\bf 34}, 450 (1981)[Pis'ma Zh. Eksp. Teor. Fiz. {\bf 34}, 471 (1981)].
 
\bibitem{dorokhov} 
O. N. Dorokhov,  JETP Lett, {\bf 36}, 318, (1982) [Pis'ma Zh. Eksp. Teor. Fiz. {\bf 36}, 259 (1982)].

\bibitem{gopar-molina} V.  A. Gopar and R.  A. Molina, Phys. Rev. B {\bf 81}, 195415 (2010).

\bibitem{fernando_gopar} F.  Falceto and V.  A. Gopar, Europhys. Lett., {\bf 92}, 57014 (2010).

\bibitem{elias} I.  Amanatidis, I.  Kleftogiannis, F. Falceto, V.  A.  Gopar, Phys. Rev. B {\bf 85}, 235450 (2012).

\bibitem{alpha1D} Notice that for 1D  lattices $\alpha = 1/2$ at the band center  (Ref. \cite{soukoulis}).

\bibitem{brouwer}P.  W. Brouwer, C. Mudry, B. D. Simons, and A. Altland , Phys. Rev. Lett.,  {\bf 81}, 862 (1998).

\bibitem{steiner} M. Steiner, Y. Chen, M. Fabrizio, and A. O. Gogolin, Phys. Rev. B {\bf 59} 14848 (1999).

\bibitem{son} Young-Woo Son, M. L. Cohen, and S. G. Louie, Phys. Rev. Lett. {\bf 97}, 216803 (2006).

\bibitem{gunlycke} D. Gunlycke and C. T. White, Phys. Rev. B {\bf 77}, 115116 (2008).

\bibitem{stone-wales} A. J. Stone and D. J. Wales, Chem. Phys. Lett. {\bf 128}, 501 (1986). 

\bibitem{ma} J. Ma, D.  Alf\`e, A. Michaelides, and E.  Wang, Phys. Rev. B {\bf 80}, 033407 (2009) 

\bibitem{ibolya} I. Zsoldos, Nanotechnol. Sci. Appl. {\bf 3}, 101 (2010).

\bibitem{valeri} V.  N. Kotov, B.  Uchoa, V.  M. Pereira, Rev. Mod. Phys. {\bf 84}, 1067 (2012).

\bibitem{araki} Y. Araki, Phys. Rev. B {\bf 84}, 113402 (2011).

\bibitem{lijie} L. Ci, Z. Xu , L. Wang , W.  Gao , F.  Ding , K. F. Kelly , B.  I. Yakobson, and P.  M.  Ajayan, Nano. Res. {\bf 1}, 116 (2008).

\bibitem{abrikosov} A. A. Abrikosov, Solid State Commun. {\bf 37}, 997 (1981). 

\end{thebibliography}
\end{document}